\documentclass[aps,twocolumn,showpacs]{revtex4}
\usepackage{epsf}
\begin{document}
\title{Conformational Transitions of Non-Grafted Polymers Near an Adsorbing Substrate}
\author{Michael Bachmann}
\email[E-mail: ]{Michael.Bachmann@itp.uni-leipzig.de}
\author{Wolfhard Janke}
\email[E-mail: ]{Wolfhard.Janke@itp.uni-leipzig.de}
\homepage[\\ Homepage: ]{http://www.physik.uni-leipzig.de/CQT}
\affiliation{Institut f\"ur Theoretische Physik, Universit\"at Leipzig,
Augustusplatz 10/11, D-04109 Leipzig, Germany}
\begin{abstract}
We have performed multicanonical chain-growth simulations 
of a polymer interacting with an adsorbing surface. The polymer,
which is not explicitly anchored at the surface, experiences a 
hierarchy of phase transitions between conformations
binding and non-binding with the substrate. We discuss the phase diagram
in the temperature--solubility plane and highlight the transition ``path''
through the free-energy landscape.
\end{abstract}
\pacs{05.10.-a, 87.15.Aa, 87.15.Cc}
\maketitle
The recent developments in single molecule experiments at the nanometer
scale, e.g., by means of atomic force microscopy (AFM)~\cite{afm1} and optical
tweezers~\cite{ot1}, allow now for a more detailed exploration of structural properties
of polymers in the vicinity of adsorbing substrates. The possibility to perform
such studies is of essential biological and technological significance. From
the biological point of view the understanding of the binding and docking mechanisms of 
proteins at cell membranes is important for the reconstruction of biological
cell processes. Similarly, specificity of peptides and binding affinity
to selected substrates could be of great importance for future electronic nanoscale
circuits and pattern recognition devices. Since single-molecule experiments
allow monitoring of polymer chains adsorbed at surfaces, the investigation of 
structural deformations of the polymer shape near substrates is a central 
aspect of experimental, computational, and theoretical studies~\cite{gray}. 

In computer simulations and analytical approaches, typically, one end of the
polymer is anchored at a flat substrate and the influence of 
adhesion and steric hindrance~\cite{grassberger1,vrbova,singh,causo1,prellberg1,huang1}, 
pulling forces~\cite{celestini1,prellberg2} or external 
fields~\cite{frey1} on the shape of the polymer is considered. The question how 
a flexible substrate, e.g., a cell membrane, bends as a reaction of a grafted polymer,
was, for example, addressed in Ref.~\cite{lipowsky1}. Proteins exhibit a strong
specificity as the affinity of peptides to adsorb at surfaces depends on the amino 
acid sequence, solvent properties, and substrate shape. This was experimentally and numerically studied, e.g.,
for peptide-metal~\cite{brown1,schulten1} and peptide-semiconductor~\cite{whaley1,goede1} interfaces.
Binding/folding and docking properties of lattice heteropolymers at an adsorbing
surface were subject of a recent numerical study~\cite{irbaeck1}. 
    
In this work we investigate in detail the
temperature and solubility dependence of adsorption properties for a polymer which 
is {\em not fixed} at the surface of the substrate with one of its ends.
This model was inspired by the experimental setup used in Refs.~\cite{whaley1,goede1},
where the peptides are initially freely moving in solution before adsorption.
Therefore, there are two main differences in comparison with studies of polymers explicitly
grafted at the substrate: First, the chain can completely desorb from the substrate 
allowing for the investigation of the binding/unbinding transition. Second,
adsorbed conformations are possible, where none of the two polymer ends is in contact with 
the surface.
 
We use a lattice model~\cite{vrbova} for a polymer near an adsorbing
substrate where nearest-neighbor contacts between monomers (nonadjacent along the chain) and 
contacts between monomers and the substrate are assigned different energy scales. The
polymer energy is given by
\begin{equation}
\label{model}
E=-\varepsilon_s n_s-\varepsilon_m n_m,
\end{equation}
where $n_{s,m}$ are the numbers of contacts with the surface and between the monomers, respectively.
The associated energy scales are set to $\varepsilon_s=1$ and $\varepsilon_m=s$,
for convenience. The solvent parameter $s$ takes account of the goodness of the implicit 
solvent and rates the two energy scales. As a first guess, $s>0$ for a poor solvent, where globular
conformations of the polymer are preferred. For $s=1$, adsorption and collapse are
equally attractive for the polymer. Negative values of $s$ refer to good solubility and 
stretched conformations dominate.
The partition sum per surface area $A$ for the polymer with the described properties at inverse
temperature $\beta=1/k_BT$ ($k_B\equiv 1$ in the following) can be written as
\begin{equation}
\label{partsum}
Z(\beta,s)/A=\sum\limits_{n_s,n_m}g_{n_s n_m}e^{\beta(n_s+s n_m)},
\end{equation} 
where 
$g_{n_s,n_m} = \delta_{n_s 0}\,g^{\rm u}_{n_m}+(1-\delta_{n_s 0})g^{\rm b}_{n_s n_m}$
is the contact density. In this decomposition, $g^{\rm u}_{n_m}$ stands for the density
of unbound conformations, whereas
$g^{\rm b}_{n_s n_m}$ is the density of surface and intrinsic contacts of all
conformations bound to the substrate. Since the number of unbound conformations in the 
half-space accessible to the polymer 
is unrestricted, $g^{\rm u}_{n_m}$ formally diverges.
%with the distance $z$ from the substrate. 
For regularization, we introduce  
an impenetrable (but neutral, i.e., non-adhesive) wall at a sufficiently large distance
$z_w$ from the substrate, in order to keep its influence on the unbound polymer small.
\begin{figure}
\centerline{\epsfxsize=8.8cm \epsfbox{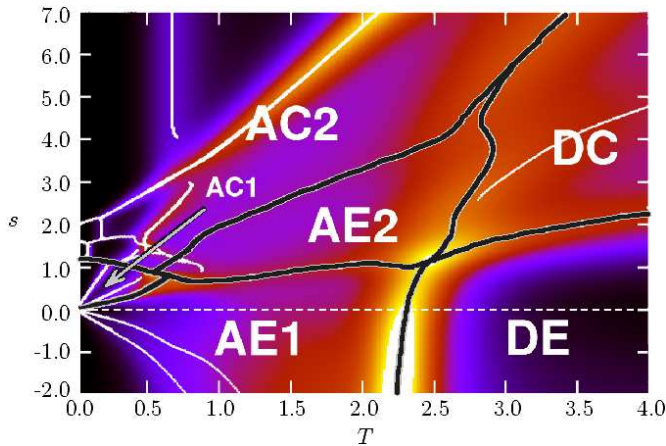}}
\caption{\label{pd} (Color online) Pseudo-phase diagram of a polymer with 100 monomers obtained from the
specific heat $C_V$ as a function of temperature $T$ and solubility parameter $s$. The 
white and black drawn lines indicate the ridges of the specific heat profile (see text). 
The dashed line separates
the regions of poor and bad solvent ($s>0$) from good solvent ($s<0$).} 
\end{figure}

We apply the powerful multicanonical chain-growth algorithm, originally introduced for the
simulation of lattice proteins~\cite{bj1}, in order to simulate the contact
density $g_{n_s n_m}$ directly. With this algorithm,
which sets up on PERM chain-growth~\cite{grass2}, the contact distributions
are flattened in a recursive way. This enables optimal sampling of the conformational
space, and all energetic quantities such as, e.g., the specific heat are obtained
by reweighting the density of contacts with respect to temperature and
solubility. The main advantage is that the whole phase diagram can in principle be 
constructed within a {\em single} simulation~\cite{prellberg3}. This method is the key for 
unravelling the detailed structure of
the phase diagram, in particular, at low temperatures, where most importance sampling 
Monte Carlo algorithms run into difficulties.
In order to break correlations being inherent
in the chain-growth process, we averaged over independent simulations (including the
determination of the multicanonical weights), and a total 
statistics of more than $10^9$ chains was accumulated in the production runs for
a homopolymer with 100 monomers. For confirmation,
we also investigated polymers with up to 200 monomers~\cite{bj3}.
 
In Fig.~\ref{pd} we show the pseudo-phase diagram of the 100mer near an attractive
substrate and the steric wall in a distance $z_w=200$ from the substrate. 
The color codes the height of the specific heat $C_V$ as the function of the 
temperature $T$ and the
solvent parameter $s$; the brighter the larger the value of $C_V$. The white
and black lines emphasise the maxima of the specific heat which shall serve as 
an orientation for the phase boundaries. While the white lines indicate pseudo-transitions
being specific for the 100mer, lines drawn in black separate regions which
are expected to be phases in the strict thermodynamic sense.  
The precise locations of transition lines in the thermodynamic limit $N\to \infty$ 
will, however, differ from the position for the finite-length system under study.
We distinguish six thermodynamic phases, four for the adsorbed (AC1, AC2, AE1, AE2)
and two for the desorbed (DC, DE) polymer~\cite{intersection}. In the adsorbed-collapsed phase 
AC1, all monomers are in contact
with the substrate and the two-dimensional (single-layer) conformation (``film'') is 
very compact. The transition from AC1 to AC2 is the layering phase transition from single
to double-layer conformations. The white transition lines within AC2 indicate pseudo-transitions
to compact conformations with more than two layers. These transitions are expected 
to disappear in the thermodynamic limit~\cite{prellberg1}. 
The transition line between AC1
and the adsorbed-expanded phase AE1 is the two-dimensional $\Theta$ collapse. It separates 
the compact single-layered conformations in AC1 from the dissolved, but still basically two-dimensional
conformations. White lines in AE1 indicate conformational transitions to unstructured conformations
extending partially into the third dimension. The substrate-contacting layer is dissolved and although
several layers can form, no explicit layering transitions are observed in this region.
In contrast to AE1, the conformations dominating phase AE2 possess a very compact
surface layer but less compact upper layers.
As in AE1, the formation of higher-order layers is not accompanied with noticeable
conformational transitions. The difference between AE1 and AE2 becomes more
apparent when approaching the unbinding transition line to the phases DE and DC, where the 
polymer has completely desorbed from the substrate: In the desorbed-expanded phase DE
random-coil conformations dominate, while in the desorbed-collapsed phase DC globular
conformations are favored. Phases DE and DC are separated by the transition line indicating 
the three-dimensional $\Theta$ collapse.

The main difference to a polymer which is {\em explicitly} anchored at the substrate with one of its 
ends, is the occurrence of the strong binding/unbinding transition between the A
and D phases. In the D phases, the polymer can move freely within the cavity, restricted
only by the presence of the two impenetrable walls. This transition influences, however, also
the conformational behavior of the polymer in the phases AE1 and AE2, the latter 
not being present for the anchored polymer. In fact, phases AC2, AE2, and DC lie within the 
DC/SAG (surface-attached globule) regime of the anchored polymer~\cite{singh,prellberg1}, whose precise
phase structure is not yet completely clarified. The phases AE1, AC1, and AC2 approximately coincide 
in the two systems for low temperatures, although the system in our study has more entropic
freedom since also adsorbed conformations are possible, where none of the ends is anchored at 
the substrate. This may have consequences for the location of transition lines. 

\begin{figure}
\centerline{\epsfxsize=8.8cm \epsfbox{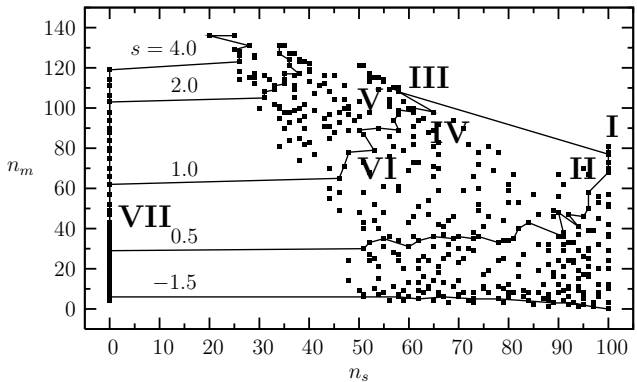}}
\caption{\label{paths} Map of free energy minima identified 
%for the parameter sets $T\in[0,4.0]$ and $s\in[-2.0,4.0]$
in the space of contact numbers $n_s$ with the substrate
and $n_m$ between monomers. Also shown are exemplified ``paths'' through the 
free-energy landscape for different fixed solvent parameters $s$. The labels
I to VII refer to the pseudo phases in the case $s=1$, described in detail 
in the text and in Table~\ref{tabPath}.
%ranging from $-1.5$ to $4$, we follow for increasing temperatures the order 
%of free energy minima taken by the ensemble starting from $T=0$.  
The lines are only guides to the eye.
}
\end{figure}
\begin{figure}[b]
\centerline{\epsfxsize=8.8cm \epsfbox{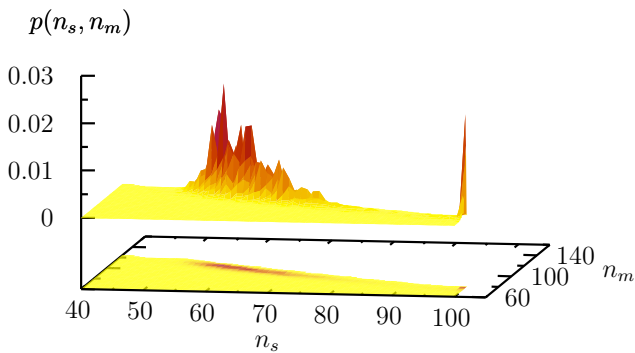}}
\caption{\label{dist} %
Probability distribution $p(n_s,n_m)$ for the 100mer in solvent with $s=1$ at
$T=0.49$, where the polymer experiences the layering transition from single
to double layer.
}
\end{figure}

Apart from the thermodynamic transitions in the traditional meaning 
we also see pseudo-transitions being specific to the chosen number of monomers
(e.g., a reorientation transition from the $5\times 5\times 4$ cube with 25
surface contacts to the rotated cube exhibiting only 20 contacts with the substrate,
which occurs within phase AC2 at $T\approx 0.7$). We are convinced that 
growing experimental capabilities will allow the 
observation of these effects also for short synthetic or naturally occurring polymers 
(e.g., peptides). Details of the different conformational phases 
will be reported elsewhere~\cite{bj3}. 

Thermodynamically, the conformations dominating a certain phase correspond
to the minimum of a suitably coarse-grained free energy depending on a few characteristic
observables of the system. The 
complexity of the free energy landscape and its dependence on external parameters
such as temperature or solvent strongly influences the kinetics of phase transitions.
For the polymer near an adsorbing surface we choose the numbers of monomer-substrate
contacts, $n_s$, and those between monomers, $n_m$, as system observables. 
According to Eq.~(\ref{partsum}), the probability
for a polymer conformation is given by 
$p(n_s,n_m)\propto g_{n_s n_m}\exp([n_s+s n_m]/T)$ and 
the contact free energy reads as
\begin{equation}
\label{free}
F_{s,T}(n_s,n_m) = -T\ln\,p(n_s,n_m),
\end{equation}   
where the temperature $T$ and the solubility $s$ are fixed external parameters.
\begin{table}
\caption{\label{tabPath} %
``Path'' through the landscape of free energy minima for a 100mer
in solvent with solubility $s=1$ with increasing temperature. 
Monomers in contact with the adsorbing substrate are shaded in light grey.
}
\begin{tabular}{cccccc}\hline\hline
\multicolumn{2}{c}{phase} & $T$ & $n_s$ & $n_m$ & typical conformations\\ \hline
 & I & $0.0$ -- $0.2$ & 100 & 81 & \parbox{3.5cm}{
\centerline{
\epsfxsize = 2.8cm \epsfbox{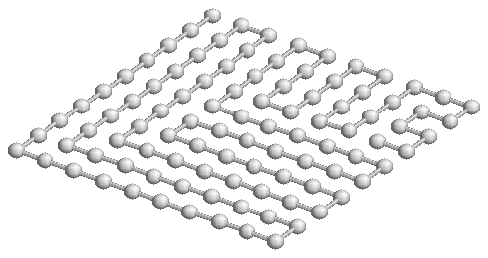}
}} \\ \cline{2-6}
\raisebox{6.2mm}[-6.2mm]{AC1} & II & $0.2$ -- $0.5$ & 100 & 77$\pm$1 &  \parbox{3.5cm}{
\centerline{
\epsfxsize = 2.8cm \epsfbox{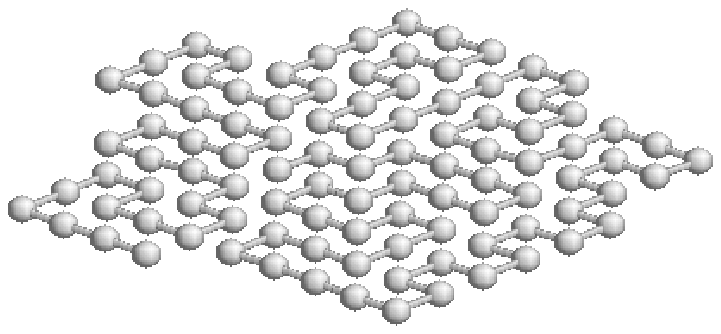}
}}\\ \hline
AC2 & III & $0.5$ -- $0.6$ & 58 & 108 & \parbox{3.0cm}{
\centerline{
\epsfxsize = 2.9cm \epsfbox{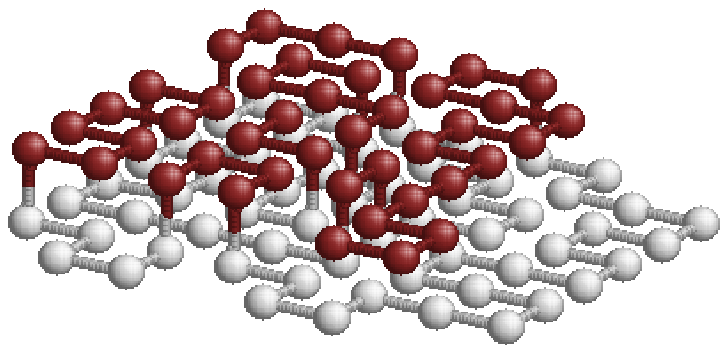}
}}\\ \hline
 & IV & $0.6$ -- $1.1$ & 61$\pm$4 & 95$\pm$5 & \parbox{3cm}{
\centerline{
\epsfxsize = 2.9cm \epsfbox{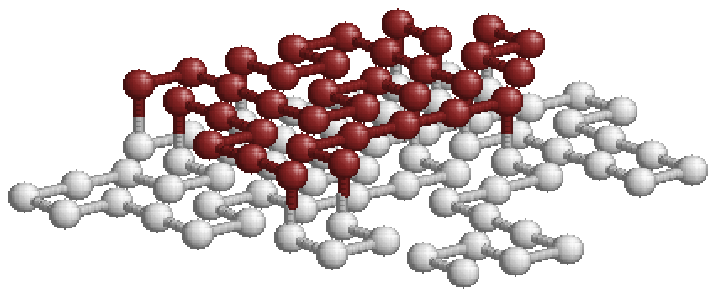}
}}\\ \cline{2-6}
\raisebox{5.8mm}[-5.8mm]{AE2} & V & $1.1$ -- $1.4$ & 53$\pm$2 & 88$\pm$2 & \parbox{3.2cm}{
\centerline{
\epsfxsize = 3.1cm \epsfbox{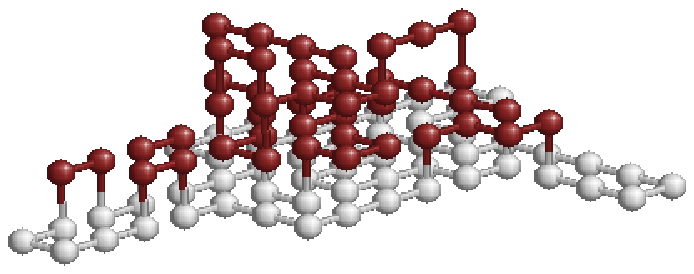}
}}\\ \hline
AE1 & VI & $1.4$ -- $2.2$ & 50$\pm$4 & 71$\pm$7 & \parbox{3.2cm}{
\centerline{
\epsfxsize = 3.1cm \epsfbox{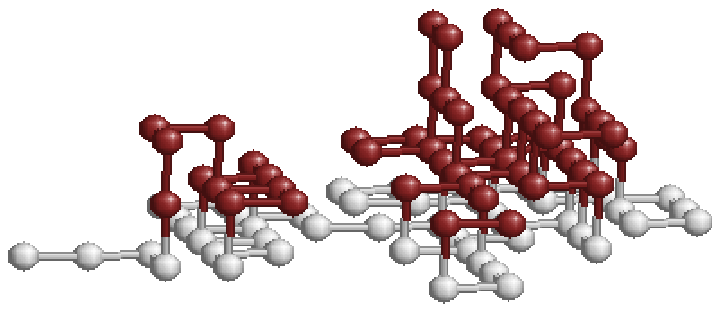}
}}\\ \hline
DE & VII & $2.2$ -- $\infty$ & 0 & $\le$62 & \parbox{2.7cm}{
\centerline{
\epsfxsize = 2.6cm \epsfbox{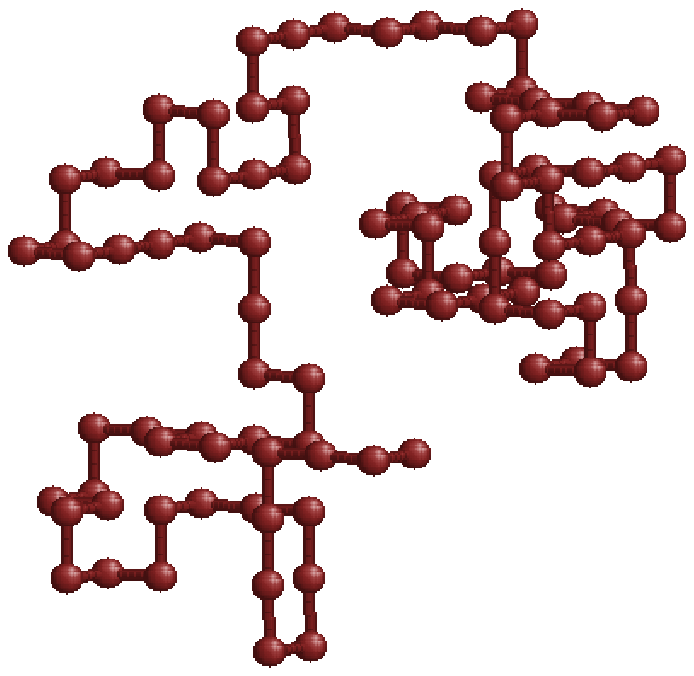}
}} \\ \hline \hline
\end{tabular}
\end{table}

In Fig.~\ref{paths} we have included all minima of the contact free energy for
the parameter set $T\in[0,4.0]$, $s\in[-2.0,4.0]$. Given a fixed solubility $s$,
the stability of a conformation with minimal free energy is connected with the
range of temperatures $\Delta T$ over which the associated free energy is actually
the global free energy minimum. We have included into Fig.~\ref{paths} 
for several fixed solubilities the ``paths'' of free energy minima
hit when increasing the temperature from $T=0$ up to $T=4$, that is moving from
right to left. 

As an example we consider the case $s=1$ for the whole region of temperatures.
In Table~\ref{tabPath} we have listed the conformational transitions the
100mer experiences by increasing the temperature.
We start at $T=0$ with the ground state which is a lamellar
two-dimensional conformation with 100 surface contacts and 81
monomer-monomer contacts. This is the maximal compact conformation
that is possible in two dimensions. Note that with {\it a conformation
characterising a phase} we mean all structures with the same
number of surface and intrinsic contacts as the minimum of the
free energy. The conformation characterising phase I is 
highly degenerate. There are in fact about $10^{14}$ (including all symmetries
except translation) different realizations of the ground state.
The ground-state conformation remains stable until $T\approx 0.2$, where the
structures become less ordered. The lamellar structure is 
dissolved, but all in all they are still 
two-dimensional and very compact. The conformational changes in the 
transition from I to II are rather local -- in contrast to the probably
actual phase transition from II to III at $T\approx 0.5$, 
where the number of surface contacts is drastically reduced to about half the
value of the ground-state conformation and thus
a second layer forms.
In Fig.~\ref{paths} this transition appears as a jump from the 
surface state $(n_s,n_m)=(100,77)$ to $(58,108)$. As can be seen in Fig.~\ref{dist},
the probability distribution $p(n_s,n_m)$ 
exhibits two distinct peaks at this temperature, 
which is interpreted as strong signal for a first-order transition.  
Entering regime IV, i.e., the adsorbed-expanded phase AE2, the dissolution of the surface-contacting  
layer begins. This process continues after passing the pseudo-transition line to
section V, where higher-order layers form. Respective bottom layer and upper layers still form
connected parts -- in contrast to phase VI (which belongs to AE1), where upper layers can break apart and
form isolated islands. 
Increasing the temperature further, we approach the second strong first-order-like transition line
and the polymer unbinds from the substrate for temperatures $T>2.2$. 
The free energy minimum jumps from the contact state $(46,65)$ to 
$(0,62)$ discontinuously (see once more Fig.~\ref{paths}). Therefore, 
the conformations occurring in phase VII do not longer prefer surface contact.
Since the thermal energy superimposes the relatively weak attraction between
the monomers at these temperatures, the 100mer in solvent with $s=1$ does not 
experience the three-dimensional $\Theta$ transition, because it is 
already in the random coil phase after the unbinding. Note that we have 
also missed the two-dimensional $\Theta$ collapse on the substrate. For this 
to happen, the quality of the solvent would have to be better (i.e., smaller values of $s$). 
From the
free-energy perspective, both collapse transitions are of second order, since 
the free-energy minima of the 100mer at $(n_s,100)$ (two-dimensional 
surface-layer conformations) and $(0,n_m)$ (three-dimensional conformations
without contact to the surface) change continuously for increasing temperature
(see Fig.~\ref{paths}).

In this work, we have qualitatively analysed the complete phase diagram 
for a polymer with 100 monomers in solvent near an adsorbing substrate. 
By means of an analysis of the global minima in the free-energy landscape 
we discussed conformational transitions in the temperature and solubility
parameter space. Two types of transitions are experienced by the polymer, phase
transitions in the thermodynamic sense and transition-type cross-over effects
which are specific to the given finite number of monomers. In the first case,
further simulations of longer polymers combined with finite-size scaling analyses 
will give estimates for the associated transition lines. 
Physically perhaps even more interesting, however, are the geometrically induced 
cross-over effects which are expected to become more and more
important as the high-resolution experimental equipment allows concrete 
measurements in the nanometer range and the design of nanoscale devices will 
take advantage of the specific properties of finite-length polymers.

We thank Karsten Goede for interesting discussions on peptide adsorption at
semiconductor surfaces.
This work is partially supported by the German-Israel-Foundation (GIF) grant 
No.\ I-653-181.14/1999. 

\begin{thebibliography}{199}
\bibitem{intersection}
The intersection of the transition lines near $T\approx 2.5$ splits in fact in two tricritical
points, as we know from first studies of polymers with up to 200 monomers~\cite{bj3}.
\end{thebibliography}
\end{document}